\documentclass[letterpaper, 10 pt, conference]{ieeeconf}  

\IEEEoverridecommandlockouts                              
\let\labelindent\relax

\usepackage{graphicx}
\usepackage{subfigure}
\usepackage{color}
\usepackage{cite}
\usepackage{amsmath}
\usepackage{enumitem}
\usepackage{comment}
\usepackage{amsthm}
\usepackage{amssymb}  
\usepackage{bbding} 
\usepackage{amsfonts}
\usepackage{tikz}  
\usetikzlibrary{fpu}
\usetikzlibrary{math}

\newtheorem{myeg}{Example}
\newtheorem{mydef}{Definition}
\newtheorem{myprob}{Problem}
\newtheorem{proposition}{Proposition}
\newtheorem{mylemma}{Lemma}
\newtheorem{theorem}{Theorem}

\title{\LARGE \bf
Sensor Deception Attacks Against Initial-State Privacy\\ in  Supervisory Control Systems
}

\author{Jingshi Yao, Xiang Yin and  Shaoyuan Li
	\thanks{This work was supported by  the National Natural Science Foundation of China (62061136004, 61803259, 61833012) and the National Key Research and Development Program of China (2018AAA0101700).}
\thanks{Jingshi Yao, Xiang Yin and Shaoyuan Li are with Department of Automation
and Key Laboratory of System Control and Information Processing,
Shanghai Jiao Tong University, Shanghai 200240, China.
        {\tt\small  \{yaojingshi,yinxiang,syli\}@sjtu.edu.cn.}}%
 }

\begin{document}

\maketitle

\thispagestyle{empty}
\pagestyle{empty}
\setlength{\abovecaptionskip}{0pt}
\setlength{\belowcaptionskip}{0pt}
\setlength{\textfloatsep}{0pt}

\begin{abstract}
This paper investigates the problem of synthesizing sensor deception attackers against privacy in the context of supervisory control of discrete-event systems (DES). We consider a DES plant controlled by a supervisor, which is subject to sensor deception attacks. Specifically, we consider an \emph{active attacker} that can tamper with the observations received by the supervisor by, e.g., hacking on the communication channel between the sensors and the supervisor. The privacy requirement of the supervisory control system is to maintain \emph{initial-state opacity}, i.e., it does not want to reveal the fact that it was initiated from a secret state during its operation. 
On the other hand, the attacker aims to deceive the supervisor, by tampering with its observations, such that initial-state opacity is violated due to incorrect control actions. In this work, we investigate from the attacker's point of view by presenting an effective approach for synthesizing  sensor attack strategies threatening the privacy of the system.  To this end, we propose the All Attack Structure (AAS) that records state estimates for both the supervisor and the attacker. This structure serves as a basis for synthesizing a sensor attack strategy. We also discuss how to simplify the synthesis complexity by leveraging the structural property of the initial-state privacy requirement. A running academic example is provided to illustrate the synthesis  procedure.
\end{abstract}

\section{Introduction}
\label{sec:intro}
With the developments of computation and communication technologies, cyber-physical system (CPS) has become the new generation of engineering systems with computation devices embedded in physical dynamics. Particularly, in cyber-physical control systems, distributed sensors and actuators  exchange information in real time. This, on the one hand, enables more flexible and intelligent control architectures, but on the other hand, makes security and privacy considerations become increasingly more important in the analysis and design of CPS \cite{car_attacks,basilio2021analysis,ma2022optimal}. 

In this paper, we study privacy issues in CPS whose high-level behaviors are abstracted as discrete-event systems (DES). Specifically, we focus on an important class of information-flow property called opacity, which captures whether or not some ``secret" of the system can be inferred by an outsider via its observation. The notion of opacity has drawn much attention in the context of DES in the past few years due to its wide applications in many engineering systems such as autonomous robots \cite{yang2020secure}, cloud computing systems \cite{zeng2021quantitative} and web services \cite{bourouis2017verification}.  
The reader is referred to  \cite{jacob_intro,wu2013comparative,balun2021comparing,wintenberg2022general} for different notions of opacity and their comparisons. In this work, we will focus on the notion of \emph{initial-state opacity} \cite{saboori2013verification}, which requires that the system can never reveal the fact that it was initiated from a secret state.  

In general, an uncontrolled open-loop system may not be opaque naturally. To ensure the consideration, \emph{supervisors} are usually used to restrict the behavior of the system such that the closed-loop system under control is opaque. In the literature on DES, many algorithms have been developed for designing opacity-enforcing supervisors; see, e.g., \cite{dubreil2010supervisory,saboori2011opacity,all-embed1,tong2018current,xie2021opacity}. 
However, all existing opacity-enforcing supervisory control systems are designed for the nominal setting in the sense that no malicious attacks exist. 
Many recent works show that, however, in networked environments, supervisors are usually subject to active and malicious attacks, which may destroy their desired closed-loop property in the nominal setting.  Particularly, the sensors of the supervisor may be compromised under \emph{sensor deception attacks} \cite{lin2021synthesis,wang2021supervisory,alves2021discrete,su2018supervisor,meira2020synthesis,meira2021synthesis} such that the supervisor may receive factitious observation tampered with by the attacker. Then based on the incorrect information, the supervisor can be misled to take incorrect actions, which may further expose its secret. 

In this paper, we investigate \emph{from the attacker's point of view}. Specifically, we assume that there already exists a well-designed supervisor for the plant such that the closed-loop system is initial-state opaque without attacks on the supervisor, i.e. when the outsider is purely passive and eavesdropping. However, we assume that an \emph{active attacker} can further deceive the supervisor by tampering with some observable events it receives.  Our objective is to synthesize such a sensor deception attacker such that  (i) it may mislead the supervisor to expose its secret initial-state, i.e., to violate initial-state opacity; and (ii) at the same time, maintain itself undetected by the supervisor.  To solve this problem, we build an information structure called the \emph{all-attack structure} (ASS) that embeds all  feasible attack strategies in it. Based on this structure, we show how to effectively extract an attacker strategy satisfying the above two requirements. 

Our work is most related to \cite{su2018supervisor,meira2020synthesis}, where the authors also investigate how to synthesize sensor deception attacks against given supervisory control systems. However, the attack objective considered therein is the violation of safety, which is defined on the actual behavior of the system. Here, we consider the violation of privacy as the attack objective, which is a property defined in the information flow. Furthermore, our synthesis algorithm leverages the structural property of initial-state opacity to mitigate the synthesis complexity. To our knowledge, the synthesis of active attackers against privacy requirements has not yet been studied in the context of supervisory control of DES. 

The remaining parts of this paper are organized as follows. 
First, we present some necessary preliminaries in Section \ref{sec:pre}. 
Then we formulate the attacker synthesis problem in Section \ref{sec:set}. 
In Section~\ref{sec:AAS}, an information-based structure called AAS is proposed, which is further simplified in  Section~\ref{sec:simplify}. Finally, we present the synthesis algorithm in Section~\ref{sec:syn} and conclude the paper in Section~\ref{sec:conclude}.

\section{Preliminary}
\label{sec:pre}
\subsection{Supervisory Control of DES}
Let $\Sigma$ be a finite set of events. A string over $\Sigma$ is a finite sequence $s=\sigma_1\cdots\sigma_n,\sigma_i\in \Sigma$; 
$|s|=n$ denotes its length;
$s^i=\sigma_1\cdots\sigma_i$ denotes the prefix of $s$ with length $i\leq n$ and $s^0=\epsilon$ is the empty string.
We denote by $\Sigma^*$ the set of all strings over $\Sigma$ including the empty string $\epsilon$. A language $L\subseteq \Sigma^*$ is a set of strings 
We define $\Sigma^\epsilon=\Sigma\cup\{\epsilon\}$.

We model a DES by a finite-state automaton 
\[
G=(X, \Sigma, \delta, X_0),
\]
where $X$ is a finite set of states; $\Sigma$ is the set of events; 
$\delta:X\times\Sigma\to X$ is the (partial) transition function;
$X_0 \subseteq X$ is the set of possible initial states.
The transition function is also extended to $\delta: X\times \Sigma^*\rightarrow X$ recursively in the usual manner. 
We define $\Delta_G(x)=\{\sigma\in \Sigma: \delta(x, \sigma)!\}$ as the set of events feasible at state $x\in X$, where $!$ means ``is defined".
The language generated from $x_0\in X_0$ is defined by  $\mathcal{L}(G,x_0)=\{s\in \Sigma^*: \delta(x_0,s)!\}$; we define
$\mathcal{L}(G)= \cup_{x_0\in X_0}\mathcal{L}(G,x_0)$. 

In the context of supervisory control, the event set $\Sigma$ is partitioned as 
\[
\Sigma=\Sigma_o\dot{\cup}\Sigma_{uo}=\Sigma_c\dot{\cup}\Sigma_{uc},
\]
where $\Sigma_{o}$ (respectively, $\Sigma_c$) is the set of observable (respectively, controllable) events, and $\Sigma_{uo}$ (respectively, $\Sigma_{uc}$) is the set of unobservable (respectively, uncontrollable) events.
We define $P:\Sigma^*\to \Sigma_o^*$ as natural projection  
that replaces unobservable events in a string by $\epsilon$, 
which  can also be extended to $P:2^{\Sigma^*}\to 2^{\Sigma_o^*}$ by $P(L)=\{ P(s)\in \Sigma_o^*: s\in L \}$.

A partial-observation supervisor is a function 
\[
S:P(\mathcal{L}(G))\to \Gamma, 
\]
where $\Gamma=\{\gamma\in2^{\Sigma}:\Sigma_{uc}\subseteq\gamma\}$ is the set of  \emph{control decisions} or control patterns.
That is, upon the occurrence of $\alpha\in \Sigma_o^*$, events in $S(\alpha)\in \Gamma$ are \emph{enabled} by the supervisor. 
We use  notation $S/G$ to represent the closed-loop system under control. 
The generated language of $S/G$ starting from $x_0\in X_0$, denoted by $\mathcal{L}(S/G,x_0)$, is  defined recursively as:
\begin{itemize}
    \item 
    $\epsilon\in \mathcal{L}(S/G,x_0)$; and 
    \item 
    for any $s\in \Sigma^*,\sigma \in \Sigma$, we have $s\sigma \in \mathcal{L}(S/G,x_0)$ iff
    $[s\in \mathcal{L}(S/G,x_0)] \wedge [s\sigma\in \mathcal{L}(G, x_0)] \wedge [\sigma\in S(P(s))]$.   
\end{itemize}
We assume without loss of generality that the supervisor does not know the initial state of the system. 
Then the language generated by the controlled system $S/G$ is defined by $\mathcal{L}(S/G)=\cup_{x_0\in X_0}\mathcal{L}(S/G,x_0)$.

Throughout the paper, we assume that supervisor $S$ is \emph{recognized} by a deterministic finite-state automaton $H=(Z, \Sigma, \xi, z_0)$ such that  
\begin{enumerate}[label=\roman*]
    \item 
    $(\forall z\in Z)(\forall \sigma\!\in\! \Sigma )[\delta(z,\sigma)\!\neq\! z\!\Rightarrow\! \sigma\!\in \!\Sigma_o]$; and 
    \item 
    $(\forall s\in \mathcal{L}(S/G))[ \Delta_{H}(\xi(z_0,s))=S(P(s))]$.  
\end{enumerate}
Then the closed-loop language can also be computed by 
$\mathcal{L}(S/G)=\mathcal{L}(G\times H)$, where $\times$ is the standard product composition of automata; see, e.g., \cite{intro}.

\begin{myeg}\upshape
Let us consider system $G$ shown in Figure~\ref{fig:1a}, where $\Sigma_o=\{b,c,d\}$  and   $\Sigma_c=\{a,c,d\}$. 
Let us consider a supervisor $S$ that disables event $c$ only when observing string $b$ and enables all events otherwise. Then supervisor $S$ can be realized by automaton 
$H$ shown in Figure \ref{fig:1b}.  
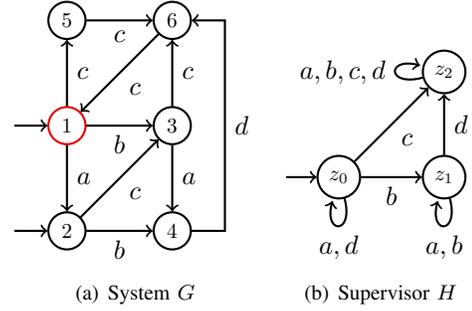
\begin{figure}
    \centering
    \subfigure[System  $G$]
    {
        \begin{tikzpicture}[state/.style={circle, draw, thick, scale=0.8}]
            \tikzmath{
                \gapx = 1.4;
                \gapy = 1.4;
            }
            \node[state, draw=red!90!black] (s1) at (0*\gapx,0*\gapy) {$1$};
            \node[state] (s2) at (0*\gapx,-1*\gapy) {$2$};
            \node[state] (s3) at (1*\gapx,0*\gapy) {$3$};
            \node[state] (s4) at (1*\gapx,-1*\gapy) {$4$};
            \node[state] (s5) at (0*\gapx,1*\gapy) {$5$};
            \node[state] (s6) at (1*\gapx,1*\gapy) {$6$};
    
            \draw[->, thick] (-\gapx/2, 0) -- (s1);
            \draw[->, thick] (s1) -- node[right]{$a$} (s2);
            \draw[->, thick] (s1) -- node[right]{$c$} (s5);
            \draw[->, thick] (s1) -- node[below]{$b$} (s3);
            \draw[->, thick] (-\gapx/2, -\gapy) -- (s2);
            \draw[->, thick] (s2) -- node[below]{$b$} (s4);
            \draw[->, thick] (s2) -- node[below right]{$c$} (s3);
            \draw[->, thick] (s3) -- node[right]{$a$} (s4);
            \draw[->, thick] (s3) -- node[right]{$c$} (s6);
            \draw[->, thick] (s4) -- (1.5*\gapx,-1*\gapy)--node[right]{$d$} (1.5*\gapx,1*\gapy) -- (s6);
            \draw[->, thick] (s5) -- node[below]{$c$} (s6);
            \draw[->, thick] (s6) -- node[below right]{$c$} (s1);
        \end{tikzpicture}
        \label{fig:1a}
    }
    \subfigure[Supervisor  $H$]
    {
        \begin{tikzpicture}[state/.style={circle, draw, thick, scale=0.8}]
            \tikzmath{
                \gapx = 1.4;
                \gapy = 1.4;
            }
    
            \node[state] (z0) at (0*\gapx,0*\gapy) {$z_0$};
            \node[state] (z2) at (1*\gapx,1*\gapy) {$z_2$};
            \node[state] (z1) at (1*\gapx,0*\gapy) {$z_1$};
    
            \draw[->, thick] (-0.5*\gapx, 0) -- (z0);
            \draw[->, thick] (z0) -- node[below right]{$c$} (z2);
            \draw[->, thick] (z0) -- node[below]{$b$} (z1);
            \draw[->, thick] (z0) edge [loop below] node{$a,d$} ();
            \draw[->, thick] (z1) -- node[right]{$d$} (z2);
            \draw[->, thick] (z1) edge [loop below] node{$a,b$} ();
            \draw[->, thick] (z2) edge [loop left] node{$a,b,c,d$} ();
        \end{tikzpicture}
        \label{fig:1b}
    }
    \caption{
    System $G$ and supervisor $H$, where $\Sigma_o=\{b,c,d\}$, $\Sigma_c=\{a,c,d\}$ and $X_{sec}=\{1\}$.}
    \label{fig:1}
\end{figure}
The closed-loop system under control $S/G$ is recognized by automaton $G\times H$ as Figure~\ref{fig:2} shows.
\begin{figure}
    \centering
    \begin{tikzpicture}[state/.style={rectangle, rounded corners, draw, thick, scale=0.8}]
            \tikzmath{
                \gapx = 1.4;
                \gapy = 1.4;
            }
            \node[state, draw=red!90!black] (s1) at (0*\gapx,0*\gapy) {$(z_0,1)$};
            \node[state] (s2) at (0*\gapx,-1*\gapy) {$(z_0,2)$};
            \node[state] (s3) at (1*\gapx,0*\gapy) {$(z_1,3)$};
            \node[state] (s4) at (1*\gapx,-1*\gapy) {$(z_1,4)$};
            \node[state] (s5) at (2*\gapx,1*\gapy) {$(z_2,5)$};
            \node[state] (s6) at (3*\gapx,1*\gapy) {$(z_2,6)$};
            \node[state] (s7) at (2*\gapx,0*\gapy) {$(z_2,1)$};
            \node[state] (s8) at (2*\gapx,-1*\gapy) {$(z_2,2)$};
            \node[state] (s9) at (3*\gapx,0*\gapy) {$(z_2,3)$};
            \node[state] (s10) at (3*\gapx,-1*\gapy) {$(z_2,4)$};
    
            \draw[->, thick] (-\gapx/2, 0) -- (s1);
            \draw[->, thick] (s1) -- node[right]{$a$} (s2);
            \draw[->, thick] (s1) -- node[below]{$b$} (s3);
            \draw[->, thick] (-\gapx/2, -\gapy) -- (s2);
            \draw[->, thick] (s2) -- node[below]{$b$} (s4);
            \draw[->, thick] (s3) -- node[right]{$a$} (s4);
            \draw[->, thick] (s1) -- node[above left]{$c$} (s5);
            \draw[->, thick] (s5) -- node[below]{$c$} (s6);
            \draw[->, thick] (s6) -- node[below right]{$c$} (s7);
            \draw[->, thick] (s7) -- node[right]{$a$} (s8);
            \draw[->, thick] (s7) -- node[right]{$c$} (s5);
            \draw[->, thick] (s7) -- node[below]{$b$} (s9);
            \draw[->, thick] (s8) -- node[below]{$b$} (s10);
            \draw[->, thick] (s8) -- node[below right]{$c$} (s9);
            \draw[->, thick] (s9) -- node[right]{$a$} (s10);
            \draw[->, thick] (s9) -- node[right]{$c$} (s6);
            \draw[->, thick] (s10) -- (3.7*\gapx,-1*\gapy) --  node[right]{$d$}(3.7*\gapx,1*\gapy)-- (s6);
            \draw[->, thick] (s2) -- (0*\gapx,-1.4*\gapy) --  node[below]{$c$}(3.5*\gapx,-1.4*\gapy)--(3.5*\gapx,0*\gapy)-- (s9);
            \draw[->, thick] (s4) -- (1*\gapx,-1.3*\gapy) --  (4*\gapx,-1.3*\gapy)--node[right]{$d$}(4*\gapx,1*\gapy)-- (s6);
        \end{tikzpicture}
    \caption{Closed-loop system $G\times H$ in Example \ref{eg:1}.}
    \label{fig:2}
\end{figure}
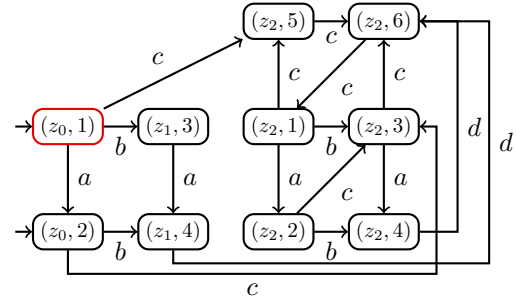 
\label{eg:1}
\end{myeg} 

Let $q\subseteq X$ be a set of states, $\gamma\subseteq \Sigma$ be a set of events and $\sigma\in \Sigma_o$ be an observable event. 
The  \emph{unobservable reach} of $q$ under event set $\gamma$ is $\textsf{UR}_\gamma(q):=\{\delta(x,s)\in X: x\in q, s \in (\Sigma_{uo}\cap\gamma)^*\}$.
The  \emph{observable reach} of $q$  upon the occurrence of  $\sigma$ is 
$\textsf{NX}_\sigma(q):=  \{\delta(x,\sigma)\in X: x\in q  \}$.
We also define $\textsf{NX}_\epsilon(q)=q$ for technical reason. 
The set of events that can be observed from $q$ under event set $\gamma$ is denoted as:
\[
\mathcal{O}(q, \gamma)=
\left\{
\sigma\in \Sigma_o\cap \gamma:  
\begin{array}{cc}
     &  
\exists x\in q, w\in (\Sigma_{uo}\cap \gamma)^*\\
     &  \text{ s.t. }\delta(x,w\sigma)!  
\end{array}\right\} 
\]

\subsection{Initial-State   and Current-State Estimates}
Given system $G$ and supervisor $S$, upon the occurrence of observable string  $\alpha\in P(\mathcal{L}(S/G))$, the supervisor can estimate the  current-state or  the initial-state of the system. 
We denote by $\mathcal{E}^C_{S/G}(\alpha)$ and $\mathcal{E}^I_{S/G}(\alpha)$, respectively, the current-state estimate and the initial-state estimate of the closed-loop system $S/G$ upon observing $\alpha$, i.e., 
\begin{align}
\mathcal{E}_{S/G}^C(\alpha)=&\left\{x\in X:\!\!
\begin{array}{cc}
& \exists x_0\in X_0, \exists s\in \mathcal{L}(S/G,x_0) \\
& \text{ s.t. } P(s)=\alpha\wedge \delta(x_0,s)=x
\end{array}\right\} \nonumber \\
\mathcal{E}_{S/G}^I(\alpha)=&\{x_0\in X_0: \exists s\in \mathcal{L}(S/G,x_0)\text{ s.t. }P(s)=\alpha\}\nonumber
\end{align} 
Using the current-state estimate,  for closed-loop system $S/G$, we have $\alpha \sigma \in P(\mathcal{L}(S/G))$ iff 
$[\alpha\in P(\mathcal{L}(S/G))]\wedge[\sigma \in \mathcal{O}(\mathcal{E}_{S/G}^C(\alpha), S(\alpha))]$.   

Note that the current-state estimate 
$\mathcal{E}_{S/G}^C(\alpha)$ 
can be computed recursively by:
\begin{itemize}
    \item 
    $\mathcal{E}_{S/G}^C(\epsilon)=\textsf{UR}_{S(\epsilon)}(X_0)$; and 
    \item 
    for any $\alpha\sigma\in P(\mathcal{L}(S/G))$, where $\sigma\in \Sigma_o$, we have
   $\mathcal{E}_{S/G}^C(\alpha\sigma)=\textsf{UR}_{S(\alpha\sigma)}(\textsf{NX}_{\sigma}(\mathcal{E}_{S/G}^C(\alpha)))$.
\end{itemize}
The initial-state estimate $\mathcal{E}_{S/G}^I(\alpha)$ can be computed through the current-state estimate over the augmented state-space \cite{shu2012detectability}. 
Specifically,  the \emph{augmented system} of $G$ is an automaton 
$\tilde{G}=(\tilde{X}, \Sigma, \tilde{\delta}, \tilde{X}_0)$,
where
$\tilde{X}\subseteq X_0\times X$ is the set of states, 
$\tilde{X}_0=\{(x_0,x_0)\in \tilde{X}: x_0\in X_0\}$ is the set of initial-states, and the transition function $\tilde{\delta}: \tilde{X}\times \Sigma\to \tilde{X}$ is defined by: for any $(x_0,x)\in \tilde{X}, \sigma\in\Sigma$, we have $\tilde{\delta}((x_0,x),\sigma)=(x_0, \delta(x,\sigma) )$. 
Then we have $\mathcal{L}(\tilde{G})=\mathcal{L}(G)$ and $\mathcal{L}(S/\tilde{G})=\mathcal{L}(S/{G})$. 
Furthermore, for any set of augmented states $\tilde{q}\subseteq \tilde{X}$, we define
$
I(\tilde{q})= \{  x_0\in X_0: (x_0,x)\in q     \}
$
as the set of its first components. 
Then for any observation $\alpha \in P(\mathcal{L}(S/G))$, we have
\begin{equation}\label{eq:ini-state}
    \mathcal{E}_{S/G}^I(\alpha) =I(     \mathcal{E}_{S/\tilde{G}}^C(\alpha)).
\end{equation} 
For the sake of clarity, we use notations $\widetilde{\textsf{NX}}_\sigma(\tilde{q})$ and $\widetilde{\textsf{UR}}_\gamma(\tilde{q})$ to denote the observable reach and the unobservable reach over the augmented state-space, respectively.

\begin{myeg}\upshape
Still, we consider system $G$ and supervisor $H$ shown in Figure~\ref{fig:1}.  Suppose that string $\alpha=b\in P(\mathcal{L}(S/G))$ is observed.  
We have 
$\mathcal{E}_{S/G}^C(\alpha)=\{3,4\}$ and $\mathcal{E}_{S/G}^I(\alpha)=\{1,2\}$. 
Soecifically, $\mathcal{E}_{S/G}^I(\alpha)$ can also be computed as follows. First, we construct the augmented system $\tilde{G}$ shown in Figure~\ref{fig:3}. 
Then we have $\mathcal{E}_{S/\tilde{G}}^C(\epsilon)=
\textsf{UR}_{S(\epsilon)}(\tilde{X}_0)
=
\{
(1,1),(1,2),(2,1),(2,2)
\}$ 
and $
\mathcal{E}_{S/ \tilde{G}}^C(b)=
\textsf{UR}_{S(b)}( \textsf{NX}_b( \mathcal{E}_{S/\tilde{G}}^C(\epsilon) )    )
=\{ (1,3), (1,4), (2,3), (2,4) \}
$.
Finally, we get
$\mathcal{E}_{S/G}^I(b) =I(     \mathcal{E}_{S/\tilde{G}}^C(b))
=\{1,2\}$.
\begin{figure}
    \centering
    \begin{tikzpicture}[state/.style={circle, draw, thick, scale=0.7}]
            \tikzmath{
                \gapx = 1.4;
                \gapy = 1.4;
            }
            \node[state, draw=red!90!black] (s1) at (0*\gapx,0*\gapy) {$1,1$};
            \node[state] (s2) at (0*\gapx,-1*\gapy) {$1,2$};
            \node[state] (s3) at (1*\gapx,0*\gapy) {$1,3$};
            \node[state] (s4) at (1*\gapx,-1*\gapy) {$1,4$};
            \node[state] (s5) at (0*\gapx,1*\gapy) {$1,5$};
            \node[state] (s6) at (1*\gapx,1*\gapy) {$1,6$};
    
            \draw[->, thick] (-\gapx/2, 0) -- (s1);
            \draw[->, thick] (s1) -- node[right]{$a$} (s2);
            \draw[->, thick] (s1) -- node[right]{$c$} (s5);
            \draw[->, thick] (s1) -- node[below]{$b$} (s3);
            \draw[->, thick] (-\gapx/2, -\gapy) -- (s2);
            \draw[->, thick] (s2) -- node[below]{$b$} (s4);
            \draw[->, thick] (s2) -- node[below right]{$c$} (s3);
            \draw[->, thick] (s3) -- node[right]{$a$} (s4);
            \draw[->, thick] (s3) -- node[right]{$c$} (s6);
            \draw[->, thick] (s4) -- (1.5*\gapx,-1*\gapy)--node[right]{$d$} (1.5*\gapx,1*\gapy) -- (s6);
            \draw[->, thick] (s5) -- node[below]{$c$} (s6);
            \draw[->, thick] (s6) -- node[below right]{$c$} (s1);
            
            \node[state, draw=red!90!black] (s1) at (2.5*\gapx,0*\gapy) {$2,1$};
            \node[state] (s2) at (2.5*\gapx,-1*\gapy) {$2,2$};
            \node[state] (s3) at (3.5*\gapx,0*\gapy) {$2,3$};
            \node[state] (s4) at (3.5*\gapx,-1*\gapy) {$2,4$};
            \node[state] (s5) at (2.5*\gapx,1*\gapy) {$2,5$};
            \node[state] (s6) at (3.5*\gapx,1*\gapy) {$2,6$};
    
            \draw[->, thick] (s1) -- node[right]{$a$} (s2);
            \draw[->, thick] (s1) -- node[right]{$c$} (s5);
            \draw[->, thick] (s1) -- node[below]{$b$} (s3);
            \draw[->, thick] (2*\gapx, -\gapy) -- (s2);
            \draw[->, thick] (s2) -- node[below]{$b$} (s4);
            \draw[->, thick] (s2) -- node[below right]{$c$} (s3);
            \draw[->, thick] (s3) -- node[right]{$a$} (s4);
            \draw[->, thick] (s3) -- node[right]{$c$} (s6);
            \draw[->, thick] (s4) -- (4*\gapx,-1*\gapy)--node[right]{$d$} (4*\gapx,1*\gapy) -- (s6);
            \draw[->, thick] (s5) -- node[below]{$c$} (s6);
            \draw[->, thick] (s6) -- node[below right]{$c$} (s1);
        \end{tikzpicture}
    \caption{Augmented system $\tilde{G}$ for system $G$ in Figure~\ref{fig:1a}}
    \label{fig:3}
\end{figure}
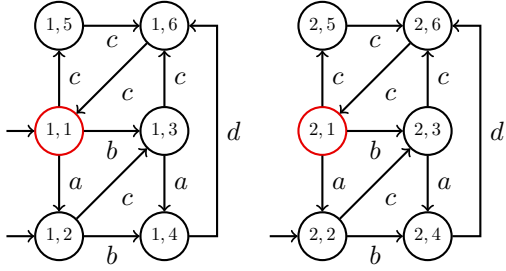
\label{eg:2}
\end{myeg}
\subsection{Initial-State Opacity}
In some cases, for example, due to insecure communications, the observation of the system  may also be available to an outsider. Here, we first consider a passive intruder (eavesdropper) with the following   capabilities:  
\begin{enumerate}[label={A}\arabic*]
    \item
    It knows both the system model $G$ and the functionality of supervisor $S$;  
    \item 
    It can also observe the occurrences of events in $\Sigma_o$.  
\end{enumerate}
Essentially, the passive intruder has exactly the same knowledge about the system as that of the supervisor. 

Furthermore, we assume that the system has some ``secret" that should not be revealed to the outside world. 
In this work, we assume that the system wants to hide the fact that it starts from some secret initial states $X_{sec}\subseteq X_0$. 
This requirement can be formalized by the notion of  \emph{initial-state opacity} defined as follows.   

\begin{mydef}[Initial-State Opacity] \upshape
Given system $G$, observable events $\Sigma_o$, supervisor $S$ and a set of secret initial states $X_{sec}\subseteq X_0$, 
the closed-loop system $S/G$ is said to be initial-state opaque w.r.t.\ $X_{sec}$ and $\Sigma_o$ if \vspace{-3pt}
\begin{align}
&(\forall x_0\in X_{sec})(\forall s\in \mathcal{L}(S/G,x_0)) \nonumber\\
&(\exists x_0'\in X_0\setminus X_{sec})( \exists t\in\mathcal{L}(S/G, x_0'))
[P(s)=P(t)]
\end{align}
or equivalently, \vspace{-3pt}
\begin{equation} 
(\forall \alpha\in P(\mathcal{L}(S/G))) [\mathcal{E}^I_{S/G}(\alpha)\not\subseteq X_{sec}].
\end{equation}{\vspace{-14pt}}
\end{mydef} 
  
In practice, the open-loop system may not be initial-state opaque without control. 
Therefore, supervisor $S$ is usually designed to guarantee that the closed-loop system $S/G$ is initial-state opaque; see, e.g., \cite{dubreil2010supervisory,saboori2011opacity,all-embed1,tong2018current,xie2021opacity} for how to synthesize such a supervisor. 
We illustrate the opacity-enforcing  supervisor by the following example.  

\begin{myeg}\label{eg:3}\upshape
We still consider system $G$ shown in Figure~\ref{fig:1a} and assume state $1$ is a secret initial-state, i.e., $X_{sec}=\{1\}$. Specifically, if the system is initiated from $1$, then when sequence $bc\in P(\mathcal{L}(G,1))\setminus P(\mathcal{L}(G,2))$ is observed, we have
$\mathcal{E}_{S/G}^I(bc)=\{1\}\subseteq X_{sec}$, i.e., the intruder knows for sure that the system was from a secret-state. 
However, under the supervision of $S$ shown in Figure~\ref{fig:1b}, the closed-loop system under control shown as Figure~\ref{fig:2} is initial-state opaque because the secret-revealing string $bc$ is on longer in the closed-loop language $\mathcal{L}(S/G)$.
\end{myeg}

\section{Active Attack Against Initial-State Opacity}
\label{sec:set}

\subsection{Motivating Example}
\label{subsec:motivating example}

In the standard problem of initial-state opacity, the intruder is assumed to be \emph{passive} in the sense that it can only ``eavesdrop" on the observations  without interfering with the observations received by the supervisor.  
In Example \ref{eg:3}, we have provided an example, where the supervisor successfully protects the initial secret of the closed-loop system from being revealed to the passive intruder. 
 
\begin{figure}
    \centering
    \subfigure[Nominal supervisory control system $S/G$.  ]{\label{model1}\includegraphics[scale=0.55]{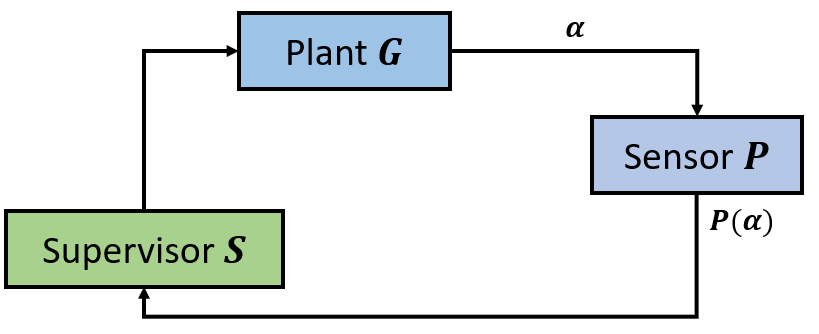}}
    \subfigure[Supervisory control system $S_A/G$ under attack.  ]{\label{model2}\includegraphics[scale=0.55]{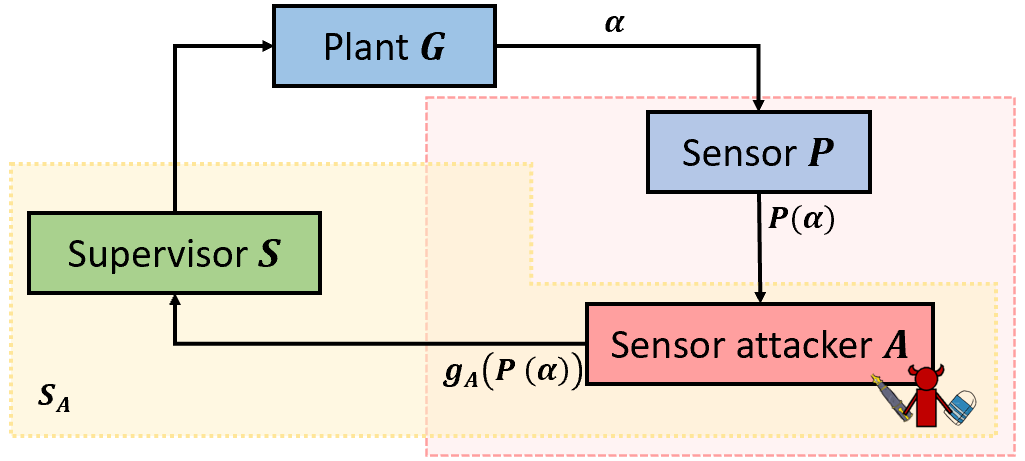}}
    \caption{Illustration of sensor attacks.}
    \label{fig:model}
\end{figure}

However, as depicted in Figure~\ref{model2}, in networked control systems, some powerful attackers may further have the capability to \emph{actively tamper} with the observations sent by the sensors in order to mislead the feedback decisions. Such an attacker model is referred to as the \emph{sensor deception attacks} \cite{su2018supervisor,meira2020synthesis}.  The following example shows that, in the case of sensor deception attacks, the attacker may actively mislead a supervisor, which is original opaque-enforcing, to expose its  secret initial state.  

\begin{myeg}\upshape
\label{eg:4}
Let us still consider system $G$ and supervisor $S$ in Figure~\ref{fig:1}. 
Now, we consider an active sensor attacker $A$ as depicted in Figure~\ref{model2}. 
Suppose that the attacker $A$ can \emph{erase} the first occurrence of observable event $b$, e.g., by  manipulating the communication channels between the sensors and the supervisor. 
Therefore,  when observable string $b\beta$ actually occurs, the supervisor will only receive $\beta$.  
Recall that supervisor $S$ simply disables event $c$ when it observes string $b$. Therefore,  under attacker $A$, string  $bc$ is still feasible since the supervisor does not know that event $b$ happens. 
Then when the attacker observes $bc$, it knows for sure that the system was initiated from secret state $1$.
\end{myeg}

\subsection{Active Attackers and Problem Formulation}
To formulate the above setting, we define $\Sigma_v \subseteq \Sigma_o$ as the set of \emph{vulnerable events} whose occurrences can potentially  be tampered with by an \emph{active attacker}. 
An active attacker (or attack strategy) is a  function 
\[
A: P(\mathcal{L}(G))\to \Sigma_o^{\epsilon},
\]
satisfying the following constraints:
\begin{itemize}
    \item 
    $A(\epsilon)=\epsilon$; 
    \item 
    for any $\alpha\sigma\in P(\mathcal{L}(G))$, 
$A(\alpha\sigma)\in \left\{
\begin{aligned}
&\{\sigma\}        &\text{if }& \sigma\notin \Sigma_v\\
&\ \Sigma_v^\epsilon &\text{if }& \sigma\in\Sigma_{v}
\end{aligned}
\right..$   
\end{itemize}
That is, upon the occurrence of each new observable   event $\sigma\in \Sigma_o$, if it is a vulnerable event, then 
the intruder may choose to either erase this event or replace it with another event in $\Sigma_v$.
An active attacker $A$ essentially induces a modification mapping for observable strings
$g_A: P(\mathcal{L}(G)) \to \Sigma_o^*$,
which is defined recursively as: 
\begin{itemize}
    \item 
    $g_A(\epsilon)=A(\epsilon)$;   and  
    \item 
       $\forall \alpha\sigma\in P(\mathcal{L}(G)): g_A(\alpha\sigma)=g_A(\alpha)A(\alpha\sigma)$.
\end{itemize}

Therefore, upon the occurrence of   $s\in \mathcal{L}(G)$,  
the supervisor will observe $g_A(P(s))$ and issue decision   $S(g_A(P(s)))$. 
Essentially, system $G$ can be regarded as being controlled by a new ``supervisor" $S_A= S\circ g_A$,
where $\circ$ is the standard function composition.
Then the language generated by the controlled system under attack is given by $\mathcal{L}( S_A/G)$.

Before the supervisor detect the presence of the attacker, 
it will estimate the state based on the doctored observation in original closed-loop system $S/G$ since it is not aware of the attacker. On the other hand,   the attacker will estimate the state based on the real observation in attacked closed-loop system $S_A/G$. In summary, upon the occurrence of actual observation $\alpha\in P(\mathcal{L}(S_A/G))$,
\begin{itemize}
    \item 
    from the supervisor's point of view, the current-state estimate and the initial-state estimate are, respectively,  $\mathcal{E}^C_{S/G}(g_A(\alpha))$ and $\mathcal{E}^I_{S/G}(g_A(\alpha))$; 
    \item 
    from the attacker's point of view, the current-state estimate and the initial-state estimate are, respectively,  
    $\mathcal{E}^C_{S_A/G}( \alpha )$ and $\mathcal{E}^I_{S_A/G}( \alpha )$.
\end{itemize}

Note that, for actual observation $\alpha\in P(\mathcal{L}(S_A/G))$, if the attacker changes it to observation $g_A(\alpha) \notin P(\mathcal{L}(S/G))$, or equivalently, 
$\mathcal{E}^C_{S/G}(g_A(\alpha))=\mathcal{E}^I_{S/G}(g_A(\alpha))=\emptyset $, 
then the supervisor will detect the presence of the attacker and then, it may upgrade the security level or take further actions to protect the system from being attacked. Formally, we say attack $A$ is \emph{stealthy} along sequence $\alpha\in P(\mathcal{L}(S_A/G))$   if
\begin{equation}\label{eq:cond1}
    \mathcal{E}^C_{S/G}(g_A(\alpha))\neq \emptyset.
\end{equation}
In this work, we investigate from the attacker's point of view.  
The objective of the attacker is to break the privacy guarantee of the system, in the sense that the initial-secret (IS) may possibly be revealed under attack. 

\begin{myprob}[{IS-Detectable Attacker Synthesis Problem}]\label{prob:main}\upshape
Given  system $G$, observable events $\Sigma_o\subseteq \Sigma$, supervisor $S$, secret initial-states $X_{sec}\subseteq X_0$ 
and vulnerable events $\Sigma_v\subseteq \Sigma_o$, synthesize an attacker $A: P(\mathcal{L}(G))\to \Sigma_o^\epsilon$ such that $\exists \alpha \sigma  \in  \mathcal{L}(S_A/G)$ satisfying:
\begin{enumerate}
    \item $A$ is stealthy along $\alpha$;
    \item $\mathcal{E}_{S_A/G}^I( \alpha \sigma )\subseteq X_{sec}$.
\end{enumerate}
If the attacker exists, we said the system $S/G$ to be IS-attackable, and the attacker $A$ to be IS-detectable.
\label{prob1}
\end{myprob}

Intuitively,  in the above formulation, $\alpha\sigma$ is a string along which the attacker can detect the secret initial state, and the supervisor should not be aware of its presence before it successfully does so.  
Here, we only require the \emph{existence} of such an attack string since we consider synthesizing an attacker that can \emph{potentially threaten} the privacy of the system. As a stronger requirement, one may also require that the attacker can \emph{always} detect the secret; this problem, however,  is beyond scope of this work.

\section{The General All Attack Structures}
\label{sec:AAS}
In this section, we define the All Attack Structure (AAS) that embeds all possible attacker's strategies in it with a suitably chosen state-space, which, therefore, can be served as the basis for our synthesis problem.  

We denote by $\hat{\Sigma}^\epsilon=\{\hat{\sigma}:\sigma\in \Sigma\}\cup \{\hat{\epsilon}\}$ the events modified by the attacker in order to distinguish from the actually events generated by the plant. 
Note that the supervisor cannot distinguish event $\sigma$ and $\hat{\sigma}$.  For every observable event $\sigma\in \Sigma_o$, we define the action space of the attacker as 
\[
\mathcal{V}(\sigma) = \left\{
\begin{array}{cl}
    \{\hat{\sigma}:\sigma\in \Sigma_v\} \cup\{\hat{\epsilon}\}      &\text{ if } \sigma\in \Sigma_v\\
    \{\hat{\sigma}\}      &\text{ if } \sigma\notin\Sigma_{v}
\end{array}
\right.. 
\]

\begin{mydef}[{All Attack  Structure}]
\label{deg_AAS}\upshape
Given system $G=(X,\Sigma,\delta,X_0)$, observable events $\Sigma_o\subseteq \Sigma$, supervisor $S$ realized by automaton $H=(Z,\Sigma,\xi,z_0)$   and vulnerable events $\Sigma_v\subseteq \Sigma_o$,  the All Attack  Structure $M$ is a new finite-state automaton
\begin{equation}
    M=(Q, \Sigma_M, f , q_0),
\end{equation} 
where  
\begin{itemize}[leftmargin=*]
\item 
   $Q=Q_e\dot{\cup} Q_a$ such that $Q_e\subseteq 2^{X}\times 2^{\tilde{X}}\times (Z\cup\{z_{\textsf{att}}\})$ is the set of \emph{environment states}, 
   where $z_{\textsf{att}}$ is a new state meaning that the supervisor has realize the presence of the attacker and
   $Q_a\subseteq 2^{X}\times 2^{\tilde{X}}\times Z\times \Sigma_o$ is the set of \emph{attack states}; 
\item 
   $q_0=( X_0, \tilde{X}_0, z_0)\in Q_e$ is the initial state;  
\item 
   $\Sigma_M=\Sigma_{o}\cup \hat{\Sigma}_{o}^{\epsilon}$ is the set of events;  
\item 
  $f:Q \times \Sigma_M \to Q$ is the transition function consists of the following two types of transitions: 
  \begin{itemize}
      \item 
      $f_{ea}:Q_e\times \Sigma_o \to Q_a$ is the transition function from 
      environment states to attack states defined by: 
      for any $q_e=(q,\tilde{q},z)\in Q_e$, we have 
      \[
      \Delta_M(q_e) = \left\{
\begin{array}{cl}
 \mathcal{O}(\tilde{q},\Delta_{H}(z))    &\text{ if }  z\in Z\\
 \emptyset     &\text{ if }  z=z_{\textsf{att}}
\end{array}
\right.. 
      \]
and for  $\sigma \!\in\!\Delta_M(q_e)$, we have 
      $f_{ea}(q_e,\sigma)\!=\! (q,\tilde{q},z,\sigma)$.
      
      \item 
      $f_{ae}:Q_a\times (\hat{\Sigma}_{o}\cup\{\hat{\epsilon}\}) \to Q_e$ is the transition function from attack states to  environment states defined by: 
      for any $q_a\!=\!(q,\tilde{q},z,\sigma)$, we have 
      $\Delta_M(q_a) \!=\! \mathcal{V}(\sigma)$. 
      For each $\hat{\sigma}_a \!\in\! \Delta_M(q_a)$, we have  
$f_{ae}(q_a, \hat{\sigma}_a )\!=\! (q',\tilde{q}',z')$, where $q'\!= \!  \textsf{NX}_{\sigma_a}(  \textsf{UR}_{\Delta_H(z)}(q) )$,  $\tilde{q}'\!=\!\widetilde{\textsf{NX}}_{\sigma}(\widetilde{\textsf{UR}}_{\Delta_H(z)}(\tilde{q}))$
  and 
\[
    z'= \left\{
\begin{aligned}
&\xi(z,\sigma_a) & \text{if } & q'\neq \emptyset\\
&z_{\textsf{att}} & \text{if } & q'= \emptyset
\end{aligned}
\right..
    \] 
  \end{itemize} 
\end{itemize}
\end{mydef}

The intuition of the AAS is explained as follows. 
The AAS essentially serves as an arena for the game between the system and the attacker. The system-player plays at  each \emph{environment state} by randomly generating a feasible observation in $\mathcal{O}(\tilde{q},\Delta_H(z))$. Whenever an observation $\sigma$ occurs, the game moves to an attack state simply by ``remembering" $\sigma$. 
Note that we use a new state $z_{\textsf{att}}$ to denote that the supervisor has detected the presence of the attacker. Therefore, if the third component of the state is $z_{\textsf{att}}$, the game stops, i.e., no active event is defined. 
The attacker-player plays at each \emph{attacker states} by choosing a doctored observation.
Note that, when the attacker changes the original observation $\sigma$ to $\sigma_a$, we use notation $\hat{\sigma}_a$ with a hat to emphasize that this is a doctored observation.
For technical reason, we define $\Delta_H(z_{\textsf{att}})=\emptyset$.
 
To formally connect the AAS structure with the attacker strategies, we introduce the following concepts. First.  we denote by $\mathcal{L}_e(M)=\{\alpha\in \Sigma_M^*:f(\alpha)\in Q_e\}$ the set of \emph{extended strings} where all strings end up with an environment state. 
Consider $\alpha=\sigma_1\hat{\sigma}_{a1}\sigma_2\hat{\sigma}_{a2}\cdots \sigma_n\hat{\sigma}_{an}\in \mathcal{L}_e(M)$.
We denote by $\textsf{obs}(\alpha)=\sigma_1\sigma_2\cdots \sigma_n$ the string consists of   odd events in $\alpha$,
and denote by $\textsf{tam}(\alpha)=\sigma_{a1}\sigma_{a2}\cdots \sigma_{an}$ the string consists of  even events in $\alpha$ with ``hats" removed. 
Therefore, for $\alpha\in \mathcal{L}_e(M)$, $\textsf{obs}(\alpha)$ is the actual observation, which is observed by the attacker, 
and $\textsf{tam}(\alpha)$ is the modified observation, which is observed by the supervisor. 
The inverse $\textsf{obs}^{-1}$ is defined by:   for every observation $\alpha=\sigma_1\cdots\sigma_n\in P(\mathcal{L}(G))$, we have $\textsf{obs}^{-1}(\alpha)=\{\sigma_1\}\mathcal{V}(\sigma_1)\cdots\{\sigma_n\}\mathcal{V}(\sigma_{n})$.

On the other hand, given an attack  strategy $A$ and an actual observation $\alpha=\sigma_1\sigma_n\dots\sigma_n\in P(\mathcal{L}(S_{A} /G  ))$,   
an extended string can be uniquely specifies as 
$
\alpha_{A}
=\sigma_1\hat{\sigma}_{a1}\sigma_2\hat{\sigma}_{a2}\cdots \sigma_n\hat{\sigma}_{an}
$
where 
$
\sigma_{ai}=A(\sigma_1\sigma_2\cdots\sigma_i)
$.
Clearly, we have  
$\textsf{obs}(\alpha_{A})= \alpha$ and  $\textsf{tam}(\alpha_{A})= g_{A}(\alpha)$.

First, we show that the first and the second components of an environment state capture, respectively, the current-state estimate of the supervisor and the initial-state estimate of the attacker. 

\begin{proposition}\label{prop:1}\upshape
For any extended string $h\!\in \!\mathcal{L}_e(M)$, let  $\beta\!=\!\textsf{tam}(h)$ and $(q,\tilde{q}, z)\!=\!f(h)$, we have
$\mathcal{E}_{S/G}^C(\beta)\!=\!\textsf{UR}_{\Delta_H(z)}(q)$.
\end{proposition}
\begin{pf}(Sketch)
The first component in $M$ is updated by an alternating sequence of 
$\textsf{UR}(\cdot)$ and $\textsf{NX}(\cdot)$ operators using the doctored observation, which is just the computation of the current-state estimate of the supervisor.  
\hfill \qed
\end{pf}

\begin{proposition}\label{prop:2}\upshape
For any extended string $h\!\in\! \mathcal{L}_e(M)$, let  $\alpha\!=\!\textsf{obs}(h), \beta\!=\!\textsf{tam}(h)$ and $(q,\tilde{q}, z)=f(h)$, 
for any attacker $A$ consistent with $h$, i.e., $g_A(\alpha)=\beta$, we have $\mathcal{E}_{S_A/G}^I(\alpha)=I(\tilde{q})$.
\end{proposition}
\begin{pf}(Sketch)
The second component  in $M$ is updated by an alternating sequence of 
$\widetilde{\textsf{UR}}(\cdot)$ and $\widetilde{\textsf{NX}}(\cdot)$ operators over the \emph{augmented} state space using the original observation,  which is essentially the computation of the augmented current-state estimate of the attacker.  
According to Eq.\ \eqref{eq:ini-state}, its first part is   the initial-state estimate of the attacker.  
\hfill \qed
\end{pf}

Also, we denote by $Q_{\textsf{att}}=\{(q, \tilde{q}, z)\in Q_e:  z=z_{\textsf{att}}\}$ the set of \emph{attack-revealing} states. Then the following result says that the attack remains stealthy along a string if and only if it does not reach an attack-revealing state in $M$. 

\begin{mylemma}\label{lemma:1}\upshape
For any attacker $A$ and observation $\alpha\sigma \in P(\mathcal{L}(S_A/G))$  such that $A$ is stealthy along $\alpha$, let 
$q_e=(q,\tilde{q},z)=f(  (\alpha\sigma)_{A}   )$ be the environment state reached by $ (\alpha\sigma)_{A}$. 
Then $A$ is stealthy along $\alpha\sigma$ if and only if 
$q_e\notin Q_{\textsf{att}}$. 
\end{mylemma}
\begin{pf}(Sketch) 
Since $\alpha$ is stealthy, $(\alpha\sigma)_{A}$ is always well-defined in $M$. 
By Proposition~\ref{prop:1}, $q=\mathcal{E}^C_{S/G}(g_A( \alpha\sigma ))$. 
Therefore, we have $q_e\in Q_{\textsf{att}}$ iff $q=\emptyset$ iff $A$ is not stealthy along $\alpha\sigma$.  
\hfill \qed
\end{pf}


\begin{myeg} \upshape
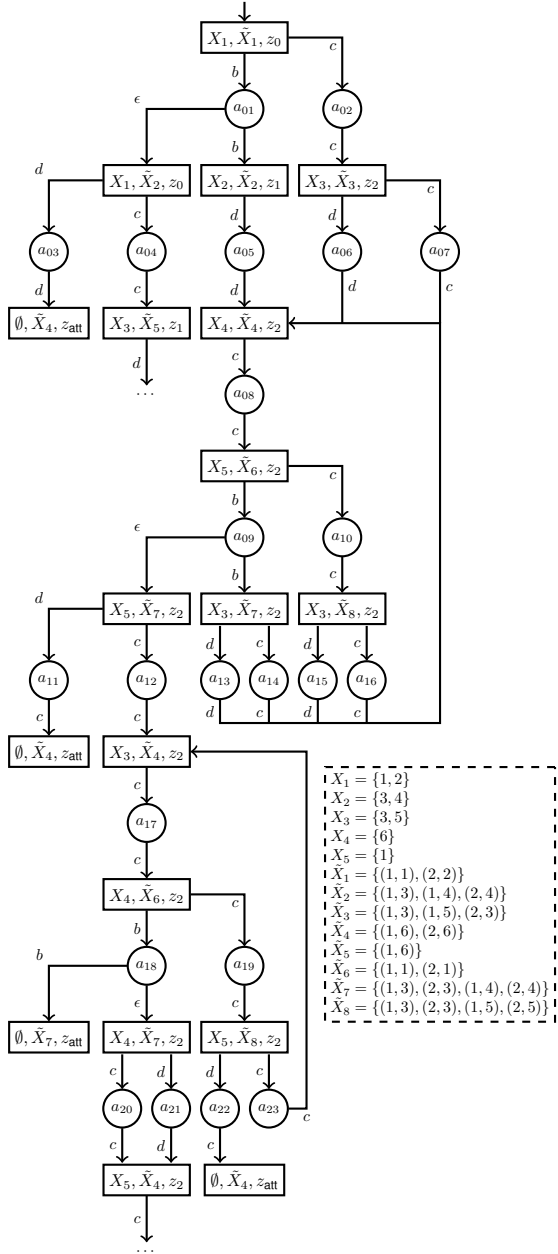
\begin{figure}
    \centering
    \begin{tikzpicture}
    [QE/.style={rectangle, draw, thick, scale=0.65}, QA/.style={circle, draw, thick, scale=0.6}]
    \tikzmath{
        \gapx = 1.3;
        \gapy = 1.9;
    }
    \node[QE] (S1) at (0*\gapx,0) {$\boldmath{X_1, \tilde{X}_1, z_0}$};
    \node[QE] (S2) at (-1*\gapx,-1*\gapy) {$X_1, \tilde{X}_2, z_0$};
    \node[QE] (S3) at (0*\gapx,-1*\gapy) {$X_2, \tilde{X}_2, z_1$};
    \node[QE] (S4) at (1*\gapx,-1*\gapy) {$X_3, \tilde{X}_3, z_2$};
    \node[QE] (S5) at (-2*\gapx,-2*\gapy) {$\emptyset, \tilde{X}_4, z_{\textsf{att}}$};
    \node[QE] (S6) at (-1*\gapx,-2*\gapy) {$X_3, \tilde{X}_5, z_1$};
    \node[QE] (S7) at (0*\gapx,-2*\gapy) {$X_4, \tilde{X}_4, z_2$};
    \node[QE] (S8) at (0*\gapx,-3*\gapy) {$X_5, \tilde{X}_6, z_2$};
    \node[QE] (S9) at (-1*\gapx,-4*\gapy) {$X_5, \tilde{X}_7, z_2$};
    \node[QE] (S10) at (0*\gapx,-4*\gapy) {$X_3, \tilde{X}_7, z_2$};
    \node[QE] (S11) at (1*\gapx,-4*\gapy) {$X_3, \tilde{X}_8, z_2$};
    \node[QE] (S12) at (-2*\gapx,-5*\gapy) {$\emptyset, \tilde{X}_4, z_{\textsf{att}}$};
    \node[QE] (S13) at (-1*\gapx,-5*\gapy) {$X_3, \tilde{X}_4, z_2$};
    \node[QE] (S14) at (-1*\gapx,-6*\gapy) {$X_4, \tilde{X}_6, z_2$};
    \node[QE] (S15) at (-2*\gapx,-7*\gapy) {$\emptyset,\tilde{X}_7, z_{\textsf{att}}$};
    \node[QE] (S16) at (-1*\gapx,-7*\gapy) {$X_4, \tilde{X}_7, z_2$};
    \node[QE] (S17) at (0*\gapx,-7*\gapy) {$X_5, \tilde{X}_8, z_2$};
    \node[QE] (S18) at (-1*\gapx,-8*\gapy) {$X_5, \tilde{X}_4, z_2$};
    \node[QE] (S19) at (0*\gapx,-8*\gapy) {$\emptyset, \tilde{X}_4, z_{\textsf{att}}$};
    
    \node[scale=0.6] (dots) at (-1*\gapx,-2*\gapy-\gapy/2) {$\cdots$};
        
    \node[QA] (A1) at (0*\gapx,0-\gapy/2) {$a_{01}$};
    \node[QA] (A2) at (1*\gapx,0*\gapy-\gapy/2) {$a_{02}$};
    \node[QA] (A3) at (-2*\gapx,-1*\gapy-\gapy/2) {$a_{03}$};
    \node[QA] (A4) at (-1*\gapx,-1*\gapy-\gapy/2) {$a_{04}$};
    \node[QA] (A5) at (0*\gapx,-1*\gapy-\gapy/2) {$a_{05}$};
    \node[QA] (A6) at (1*\gapx,-1*\gapy-\gapy/2) {$a_{06}$};
    \node[QA] (A7) at (2*\gapx,-1*\gapy-\gapy/2) {$a_{07}$};
    \node[QA] (A8) at (0*\gapx,-2*\gapy-\gapy/2) {$a_{08}$};
    \node[QA] (A9) at (0*\gapx,-3*\gapy-\gapy/2) {$a_{09}$};
    \node[QA] (A10) at (1*\gapx,-3*\gapy-\gapy/2) {$a_{10}$};
    \node[QA] (A11) at (-2*\gapx,-4*\gapy-\gapy/2) {$a_{11}$};
    \node[QA] (A12) at (-1*\gapx,-4*\gapy-\gapy/2) {$a_{12}$};
    \node[QA] (A13) at (-0.25*\gapx,-4*\gapy-\gapy/2) {$a_{13}$};
    \node[QA] (A14) at (0.25*\gapx,-4*\gapy-\gapy/2) {$a_{14}$};
    \node[QA] (A15) at (0.75*\gapx,-4*\gapy-\gapy/2) {$a_{15}$};
    \node[QA] (A16) at (1.25*\gapx,-4*\gapy-\gapy/2) {$a_{16}$};
    \node[QA] (A17) at (-1*\gapx,-5*\gapy-\gapy/2) {$a_{17}$};
    \node[QA] (A18) at (-1*\gapx,-6*\gapy-\gapy/2) {$a_{18}$};
    \node[QA] (A19) at (0*\gapx,-6*\gapy-\gapy/2) {$a_{19}$};
    \node[QA] (A20) at (-1.25*\gapx,-7*\gapy-\gapy/2) {$a_{20}$};
    \node[QA] (A21) at (-0.75*\gapx,-7*\gapy-\gapy/2) {$a_{21}$};
    \node[QA] (A22) at (-0.25*\gapx,-7*\gapy-\gapy/2) {$a_{22}$};
    \node[QA] (A23) at (0.25*\gapx,-7*\gapy-\gapy/2) {$a_{23}$};
    \node[scale=0.6] (dots2) at (-1*\gapx,-8*\gapy-\gapy/2) {$\cdots$};

    \draw[->, thick] (0*\gapx, \gapy/4) -- (S1.north);
    \draw[->, thick] (S1) -- node[left, scale=0.6]{$b$} (A1);
    \draw[->, thick] (S1) -| node[below left, scale=0.6]{$c$} (A2);
    \draw[->, thick] (A1) -| node[above left, scale=0.6]{$\epsilon$} (S2);
    \draw[->, thick] (A1) --node[left, scale=0.6]{$b$} (S3);
    \draw[->, thick] (A2) -- node[left, scale=0.6]{$c$} (S4);
    \draw[->, thick] (S2) -| node[above left, scale=0.6]{$d$} (A3);
    \draw[->, thick] (S2) -- node[left, scale=0.6]{$c$} (A4);
    \draw[->, thick] (S3) -- node[left, scale=0.6]{$d$} (A5);
    \draw[->, thick] (S4) -- node[left, scale=0.6]{$d$} (A6);
    \draw[->, thick] (S4) -| node[below left, scale=0.6]{$c$} (A7);
    \draw[->, thick] (A3) -- node[left, scale=0.6]{$d$} (S5);
    \draw[->, thick] (A4) -- node[left, scale=0.6]{$c$} (S6);
    \draw[->, thick] (A5) -- node[left, scale=0.6]{$d$} (S7);
    \draw[thick] (A6) -- node[above right, scale=0.6]{$d$} +(0,-\gapy/2);
    \draw[thick] (A7) -- node[above right, scale=0.6]{$c$} +(0,-\gapy/2);
    \draw[->, thick] (S6) -- node[left, scale=0.6]{$d$} (dots);
    \draw[->, thick] (S7) -- node[left, scale=0.6]{$c$} (A8);
    \draw[->, thick] (A8) -- node[left, scale=0.6]{$c$} (S8);
    \draw[->, thick] (S8) -- node[left, scale=0.6]{$b$} (A9);
    \draw[->, thick] (S8) -| node[below left, scale=0.6]{$c$} (A10);
    \draw[->, thick] (A9) -| node[above left, scale=0.6]{$\epsilon$} (S9);
    \draw[->, thick] (A9) -- node[left, scale=0.6]{$b$} (S10);
    \draw[->, thick] (A10) -- node[left, scale=0.6]{$c$} (S11);
    \draw[->, thick] (S9) -| node[above left, scale=0.6]{$d$} (A11);
    \draw[->, thick] (S9) -- node[left, scale=0.6]{$c$} (A12);
    \draw[->, thick] (A13)+(0,\gapy*0.38) -- node[left, scale=0.6]{$d$} (A13);
    \draw[->, thick] (A14)+(0,\gapy*0.38) -- node[left, scale=0.6]{$c$} (A14);
    \draw[->, thick] (A15)+(0,\gapy*0.38) -- node[left, scale=0.6]{$d$} (A15);
    \draw[->, thick] (A16)+(0,\gapy*0.38) -- node[left, scale=0.6]{$c$} (A16);
    \draw[->, thick] (A11) -- node[left, scale=0.6]{$c$} (S12);
    \draw[->, thick] (A12) -- node[left, scale=0.6]{$c$} (S13);
    \draw[->, thick] (A13) |- node[above left, scale=0.6]{$d$} (2*\gapx, -4.8*\gapy) |- (S7);
    \draw[thick] (A14) --  (0.25*\gapx, -4.8*\gapy)node[above left, scale=0.6]{$c$};
    \draw[thick] (A15) --  (0.75*\gapx, -4.8*\gapy) node[above left, scale=0.6]{$d$};
    \draw[thick] (A16) --  (1.25*\gapx, -4.8*\gapy) node[above left, scale=0.6]{$c$};
    \draw[->, thick] (S13) -- node[left, scale=0.6]{$c$} (A17);
    \draw[->, thick] (A17) -- node[left, scale=0.6]{$c$} (S14);
    \draw[->, thick] (S14) -- node[left, scale=0.6]{$b$} (A18);
    \draw[->, thick] (S14) -| node[below left, scale=0.6]{$c$} (A19);
    \draw[->, thick] (A18) -| node[above left, scale=0.6]{$b$} (S15);
    \draw[->, thick] (A18) -- node[left, scale=0.6]{$\epsilon$} (S16);
    \draw[->, thick] (A19) -- node[left, scale=0.6]{$c$} (S17);
    \draw[->, thick] (A20)+(0,\gapy*0.38) -- node[left, scale=0.6]{$c$} (A20);
    \draw[->, thick] (A21)+(0,\gapy*0.38) -- node[left, scale=0.6]{$d$} (A21);
    \draw[->, thick] (A22)+(0,\gapy*0.38) -- node[left, scale=0.6]{$d$} (A22);
    \draw[->, thick] (A23)+(0,\gapy*0.38) -- node[left, scale=0.6]{$c$} (A23);
    \draw[->, thick] (A20) -- node[left, scale=0.6]{$c$} +(0,-\gapy*0.38);
    \draw[->, thick] (A21) -- node[left, scale=0.6]{$d$} +(0,-\gapy*0.38);
    \draw[->, thick] (A22) -- node[left, scale=0.6]{$c$} +(0,-\gapy*0.38);
    \draw[->, thick] (A23)  -- +(0.5,0) node[below, scale=0.6]{$c$} |-  (S13);
    \draw[->, thick] (S18) -- node[left, scale=0.6]{$c$} (dots2);
    
    \node[align=left, scale=0.6, draw, thick, dashed] (words) at (2*\gapx, -6*\gapy)
        {$X_1=\{1,2\}$\\ $X_2=\{3,4\}$\\ $X_3=\{3,5\}$\\ $X_4=\{6\}$ \\ $X_5=\{1\}$ \\ 
        $\tilde{X}_1=\{(1,1),(2,2)\}$\\ $\tilde{X}_2=\{(1,3),(1,4),(2,4)\}$\\ $\tilde{X}_3=\{(1,3),(1,5),(2,3)\}$\\
        $\tilde{X}_4=\{(1,6),(2,6)\}$\\ $\tilde{X}_5=\{(1,6)\}$\\ $\tilde{X}_6=\{(1,1),(2,1)\}$\\
        $\tilde{X}_7=\{(1,3),(2,3),(1,4),(2,4)\}$\\ $\tilde{X}_8=\{(1,3),(2,3),(1,5),(2,5)\}$};
\end{tikzpicture}
    \caption{(Part of) the AAS for system $G$ and supervisor $S$ for Example \ref{eg:5}.}
    \label{fig:5}
\end{figure}
Still, let us consider   system $G$ and supervisor $S$ shown in Figure~\ref{fig:1}.  Figure~\ref{fig:5} shows part of the AAS $M$. The initial state is $(X_1, \tilde{X}_1, z_0)$, which is the tuple of the initial-state sets of $G$, $\tilde{G}$ and  $H$.  From the initial-state. the system randomly chooses an event in $\{b,c\}=\mathcal{O}_{\tilde{G}}(\tilde{X}_1, \Delta_{H}(z_0))$ to play. 
If event $b$ is chosen, the AAS will enters attacker state $a_{01}$. 
The attacker can choose an event in $\mathcal{V}(b)=\{\hat{\epsilon}, \hat{b}\}$ to play. 
If $\hat{\epsilon}$ is chosen, the supervisor observes nothing. 
Therefore, the first and the third components do not change. 
However, the attack knows that $b$ is generated and therefore, the second component is updated to $\tilde{X}_2=\widetilde{\textsf{NX}}_b(\widetilde{\textsf{UR}}_{\Delta_H(z_0)}(\tilde{X}_1))= \{(1,3),(1,4),(2,4)\}$. 
As a result, AAS enters environment state $(X_1,\tilde{X}_2,z_0)$. 
Again, the system randomly chooses an event in $\{d,c\}=\mathcal{O}_{\tilde{G}}(\tilde{X}_2, \Delta_{H}(z_0))$ to play. 
If event $d$ is chosen, the AAS will enters attacker state $a_{03}$. Since $\mathcal{V}(d)=\{\hat{d}\}$, the attacker can only choose $\hat{d}$ to play, i.e., it cannot modify the observation. 
However, in the control logic of supervisor $S$, $d\not\in \mathcal{O}(X_1, \Delta_{H}(z_0))$. 
Therefore,  the supervisor will be aware of presence of the attacker. 
As a result, AAS $M$ enters a attack-revealing state $(\emptyset, \tilde{X}_4, z_{\textsf{att}})$. 
The remaining part of the AAS structure is constructed in the same way.
\label{eg:5}
\end{myeg}
 
\section{The Simplified AAS}\label{sec:simplify}
In the previous section, we have introduced the AAS that embeds all attack strategies until their presence is revealed.   
In practice, once the attacker knows for sure that the system was or was not initiated from a secret state, the attacker-player's win or loss has been completely determined. Based on this idea, the AAS can be simplified.


\begin{mydef}[{Detected States}]\upshape
Given AAS $M$, for any environment state $q_e=(q,\tilde{q},z)\in Q_e$, we say 
$q_e$ is a
\begin{itemize}
    \item 
    \emph{positive detected state} if $I(\tilde{q})\subseteq X_{sec}$; and
    \item 
    \emph{negative detected state} if $I(\tilde{q})\cap X_{sec}=\emptyset$.
\end{itemize}
We denote by $Q_{det}^+\subseteq Q_e$ and $Q_{det}^-\subseteq Q_e$ the set of  all positive and negative detected states, respectively, and define $Q_{det} =Q_{det}^+\cup Q_{det}^-$ as the set of detected states.
\end{mydef}
 
Furthermore, we observe that, once we know for sure that the system was or was not started from secret states, we know this information forever. In terms of the AAS,  this observation is summarized by the following result. 

\begin{proposition}\label{prop:detec-state}\upshape
Let $q_e\in Q_M$ be  a positive (respectively, negative) detected state. 
Then all  environment states reachable from $q_e$ are positive (respectively, negative) detected states.
\end{proposition}

Positive/negative detected environment states essentially provide a state-based winning condition for the attacker-player. In some cases, we can also determine in advance that the attacker \emph{cannot} win the game even before it reveals itself. This is captured by the notion of undetectable states. 

\begin{mydef}[{Undetectable States}]\upshape\label{def_ud}
Given AAS $M$, for any environment state $q_e=(q,\tilde{q},z)\in Q_e$, we say 
$q_e$ is a \emph{undetectable state} if  
(i)    $I(\tilde{q})\cap X_{sec}\neq \emptyset$; and 
(ii)  for any  $ (x_0, x)\in\widetilde{\textsf{UR}}_{\Delta_H(z)}(\tilde{q})$ such that $x_0 \in X_{sec}$, there exists  $(x_0', x)\in \widetilde{\textsf{UR}}_{\Delta_H(z)}(\tilde{q})$ such that $x_0'\not\in X_{sec}$. We denote by $Q_{ud} \subseteq Q_e$ the set of all undetectable states.
\end{mydef}

Intuitively, if the system reaches an undetectable state, then it means that \emph{whenever} the system was possibly initiated from a secret state $x_0\in X_{sec}$, 
it is also possible that the system was initiated from a  non-secret state $x_0\in X_{sec}$
such that the two cases end up with the same current-state $x$ via trajectories having the same observation.  Therefore, once the system reaches  an undetectable state, we can conclude immediately that the attacker can no longer detect the secret in the future when actually the initial state is secret. 


\begin{proposition}\label{prop:sud}\upshape
Let $q_e\in Q_M$ be  an undetectable state in  AAS $M$. 
Then all environment states reachable from $q_e$ are either undetectable or negative detected states.
\end{proposition}
\begin{pf}
Let us consider  an arbitrary environment states $q_e'=(q',\tilde{q}',z')$ reachable from $q_e=(q,\tilde{q},z)$. 
If   $I(\tilde{q}')\cap X_{sec}=\emptyset$, then it is negative detected.  
If   $I(\tilde{q}')\cap X_{sec}\neq \emptyset$, 
then for any $(x_0,x')\in \tilde{q}'$ such that $x_0\in X_{sec}$, 
since  $q_e'$ is reachable from $q_e$, 
there exists $(x_0,x)\in \tilde{q}$. 
However, since $q$ is undetectable, we know that there exists $(x_0',x)\in \tilde{q}$ such that $x\not\in X_{sec}$. This further implies that $(x_0,x’) \in \tilde{q}'$. 
Therefore, we conclude that $q'$ is also undetectable. 
 
    
    

\end{pf}

Based on Propositions~\ref{prop:detec-state} and~\ref{prop:sud}, 
we know that once  we reach   a detected state in $Q_{det}$, the attacker knows for sure whether the initial state is secret, and once we reach an undetectable state in $Q_{ud}$, the attacker will never be able to discover the initial secret when the truth is secret. Therefore, there is no need to further expand the AAS from detected states and undetectable states, which leads to the simplified AAS (SAAS). Specifically,  we denote by 
\[
     M^{s}=(Q^{s}, \Sigma_M, f^{s} , q_0),
\]
the \emph{simplified AAS}, which is  
the reachable part of the AAS $M$ by removing all outgoing transitions from states 
$Q_{det}\cup Q_{ud}$. 

\begin{myeg}\upshape
For the  AAS shown in Example $\ref{eg:5}$, its SAAS is given in Figure~\ref{fig:6}. 
For environment state $(X_3,\tilde{X}_5,z_1)$,  
since $I(\tilde{X}_5)=I(\{(1,6)\})=\{1\}\subseteq X_{sec}$, it is positive detected. 
For environment state $(X_4,\tilde{X}_4,z_2)$, since $\widetilde{\textsf{UR}}_{\Delta_H(z_2)}(\tilde{X}_4)=\{(1,6),(2,6)\}$, 
where $1\in X_{sec}$ but $2\notin X_{sec}$,  it is  undetectable. 
In fact, the revealing state  $(\emptyset, \tilde{X}_4, z_{\textsf{att}})$ is also undetectable. 
Therefore, we can  remove active events defined at these states from the AAS, which gives the SAAS a much smaller state space. 
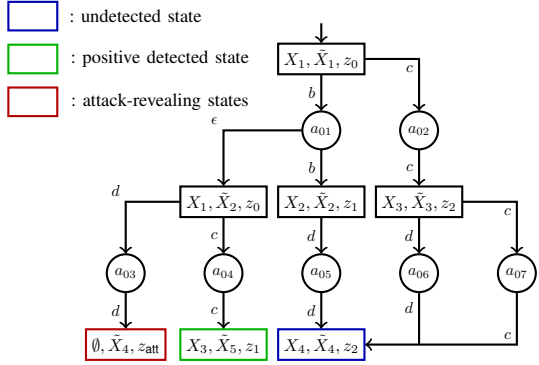
\begin{figure}
    \centering
    \begin{tikzpicture}
        [QE/.style={rectangle, draw, thick, scale=0.65}, QA/.style={circle, draw, thick, scale=0.6}]
        \tikzmath{
        \gapx = 1.3;
        \gapy = 1.9;
        }
        \node[QE] (S1) at (0*\gapx,0) {$\boldmath{X_1, \tilde{X}_1, z_0}$};
        \node[QE] (S2) at (-1*\gapx,-1*\gapy) {$X_1, \tilde{X}_2, z_0$};
        \node[QE] (S3) at (0*\gapx,-1*\gapy) {$X_2, \tilde{X}_2, z_1$};
        \node[QE] (S4) at (1*\gapx,-1*\gapy) {$X_3, \tilde{X}_3, z_2$};
        \node[QE, draw=red!70!black] (S5) at (-2*\gapx,-2*\gapy) {$\emptyset, \tilde{X}_4, z_{\textsf{att}}$};
        \node[QE, draw=green!70!black] (S6) at (-1*\gapx,-2*\gapy) {$X_3, \tilde{X}_5, z_1$};
        \node[QE, draw=blue!70!black] (S7) at (0*\gapx,-2*\gapy) {$X_4, \tilde{X}_4, z_2$};
        
        \node[QA] (A1) at (0*\gapx,0-\gapy/2) {$a_{01}$};
        \node[QA] (A2) at (1*\gapx,0*\gapy-\gapy/2) {$a_{02}$};
        \node[QA] (A3) at (-2*\gapx,-1*\gapy-\gapy/2) {$a_{03}$};
        \node[QA] (A4) at (-1*\gapx,-1*\gapy-\gapy/2) {$a_{04}$};
        \node[QA] (A5) at (0*\gapx,-1*\gapy-\gapy/2) {$a_{05}$};
        \node[QA] (A6) at (1*\gapx,-1*\gapy-\gapy/2) {$a_{06}$};
        \node[QA] (A7) at (2*\gapx,-1*\gapy-\gapy/2) {$a_{07}$};
        
        \draw[thick, blue!70!black] (-3.2*\gapx,0.4*\gapy) rectangle (-2.7*\gapx,0.2*\gapy);
        \draw[thick, green!70!black] (-3.2*\gapx,0.1*\gapy) rectangle (-2.7*\gapx,-0.1*\gapy);
        \draw[thick, red!70!black] (-3.2*\gapx,-0.2*\gapy) rectangle (-2.7*\gapx,-0.4*\gapy);
        
        \node[scale=0.7,align=left] (word1) at (-1.9*\gapx, 0.3*\gapy)
        {: undetected state};
        \node[scale=0.7,align=left] (word2) at (-1.65*\gapx, -0*\gapy)
        {: positive detected state};
        \node[scale=0.7,align=left] (word3) at (-1.65*\gapx, -0.3*\gapy)
        {: attack-revealing states};

        \draw[->, thick] (0*\gapx, \gapy/4) -- (S1.north);
        \draw[->, thick] (S1) -- node[left, scale=0.6]{$b$} (A1);
        \draw[->, thick] (S1) -| node[below left, scale=0.6]{$c$} (A2);
        \draw[->, thick] (A1) -| node[above left, scale=0.6]{$\epsilon$} (S2);
        \draw[->, thick] (A1) --node[left, scale=0.6]{$b$} (S3);
        \draw[->, thick] (A2) -- node[left, scale=0.6]{$c$} (S4);
        \draw[->, thick] (S2) -| node[above left, scale=0.6]{$d$} (A3);
        \draw[->, thick] (S2) -- node[left, scale=0.6]{$c$} (A4);
        \draw[->, thick] (S3) -- node[left, scale=0.6]{$d$} (A5);
        \draw[->, thick] (S4) -- node[left, scale=0.6]{$d$} (A6);
        \draw[->, thick] (S4) -| node[below left, scale=0.6]{$c$} (A7);
        \draw[->, thick] (A3) -- node[left, scale=0.6]{$d$} (S5);
        \draw[->, thick] (A4) -- node[left, scale=0.6]{$c$} (S6);
        \draw[->, thick] (A5) -- node[left, scale=0.6]{$d$} (S7);
        \draw[thick] (A6) -- node[above left, scale=0.6]{$d$} +(0,-\gapy/2);
        \draw[->, thick] (A7)  |-node[above left, scale=0.6]{$c$}  (S7);
    \end{tikzpicture}
    \caption{The simplified AAS of the AAS in Figure~\ref{fig:5}.}
    \label{fig:6}
\end{figure}
\end{myeg}

The  simplified AAS can be constructed by either a depth-first search or a breath-first search that starts from the initial state $q_0$, lists all feasible transitions and stops whenever a detected or undetectable state is reached. 
In the worst-case, the SAAS also contains $2^{|X|}\cdot2^{|X_0|\times |X|}\cdot|Z|$ states and 
$2^{|X|}\cdot2^{|X_0|\times |X|}\cdot|Z|\cdot|\Sigma_o|$ transitions, which are the same as the AAS. In practice, however, the SAAS can be much smaller than the AAS as illustrated by the previous example.  

\section{Synthesis of Attack Strategies}\label{sec:syn} 
In this section, we solve Problem \ref{prob1} by synthesizing a sub-system of the SAAS, called the single attack structure, that meets certain requirements. Then we prove that the existence of such a sub-system is a sufficient and necessary condition for the existence of an IS-detectable attacker.

\subsection{Single Attack  Structure}
Note that in the SAAS, at each attack state, the attacker may have multiple choices. To synthesize a deterministic attack strategy, the single attacker structure (SAS) is defined, which is a sub-system of the SAAS such that each attack state only has no more than one active event.
\begin{mydef}[{Single Attack  Structure}]\label{def_SAE}
\upshape
Let $m=(Q^m=Q_a'\cup Q_e', \Sigma_M, f^m, q_0 )$ be a sub-system of the SAAS $M^s$. 
We say $m$ is  a \emph{single attack structure}  if 
\begin{enumerate}
    \item 
    for any state $q_a\in Q_a'$, we have $|\Delta_m(q_a)|=1$;
    \item 
    for any state $q_e\in Q_e'$, we have $|\Delta_m(q_e)|=|\Delta_{M^s}(q_e)|$.  
\end{enumerate}
We denote the set of all SAS of the SAAS $M^s$ as $\mathbb{S}(M^s)$.
\end{mydef}

Intuitively, in a single attack structure, at each attack state, it only has \emph{a unique choice} of attack strategy, and at each environment state, it can reactive to all  possible system events. 
Recall that, for any observation $\alpha\in P(\mathcal{L}(G))$, $\textsf{obs}^{-1}(\alpha) \in \Sigma_M^*$ is the set of all extended strings with observation $\alpha$. 
However, in terms of SAS $m$, we know that the cardinality of
$\textsf{obs}^{-1}(\alpha)\cap \mathcal{L}_e(m)$ is always smaller than or equal to one. 
Therefore, 
we denote  by $\textsf{obs}^{-1}_m(\alpha)$ the unique extended string $m$ such that  
 $\textsf{obs}(\textsf{obs}^{-1}_m(\alpha))=\alpha$ when it exists.

Therefore, given a SAS $m\in  \mathbb{S}(M^s)$, we can uniquely an attack strategy $A_{m}$ by: 
for any observation $\alpha\sigma\in P(\mathcal{L}(G))$, 
\begin{itemize}
    \item 
    If $\textsf{obs}^{-1}_m(\alpha)$ exists,  
    then  $A_m(\alpha)={\sigma}_a$, where $\hat{\sigma}_a$ is the last event of  $\textsf{obs}^{-1}_m(\alpha)$; 
    \item 
    If $\textsf{obs}^{-1}_m(\alpha)$ does not exist, then 
     $ A_m(\alpha )=\sigma$, where $\sigma$ is the last event of $\alpha$.  
\end{itemize} 
We call such   strategy $A_m$ the SAS $m$ \emph{induced strategy}. 
We can easily show, by induction,  that $\alpha_{A_m}= \textsf{obs}^{-1}_m( \alpha)$ for any observation $\alpha$. 
Therefore, by understanding how a SAS can ``encode" an attacker strategy, hereafter, we will focus on finding a SAS instead  of finding an attacker strategy.  




\subsection{Synthesis of Attacker Strategy}
Now, we show how to synthesize an attack strategy, which is encoded as a SAS, such that Problem~\ref{prob:main} is solved. 
The following result shows that, to order to ensure an attack sequence in its induce strategy, the necessary and sufficient condition is to have a positive detected state in the SAS.  
\begin{theorem}\label{them:state-stat}\upshape
Let $m\in \mathbb{S}( M^s)$ be a SAS. 
Then $A_m$  solves Problem~\ref{prob:main}, i.e., $A_m$ is IS-detectable,  iff $Q^m\cap Q_{det}^+\neq \emptyset$. 
\end{theorem} 
\begin{pf}To see the sufficiency, suppose that $m$ contains a reached positive detected state. 
Let  $\alpha\in \mathcal{L}_e(m)$ be an extended string such that $f^m(\alpha)\in Q_{det}^+$. 
Let $\beta=\textsf{obs}(\alpha)$. Clearly, $A_m$ is stealthy along  $\beta^{|\beta|-1}$ since the sequence is terminated once it reached a revealing state. 
Furthermore, since $f^m(\alpha)=(q,\tilde{q},z)\in Q_{det}^+$, we have $\mathcal{E}_{S_{A_m}/G}^I(\alpha)=I(\tilde{q})\subseteq X_{sec}$. 
The necessity is similar, since $Q^m\cap Q_{det}^+ = \emptyset$, then for all $\alpha\in \mathcal{L}(S_{A_m}/G)$, it either is not stealthy or can not detect the initial secret using the same argument.
\hfill\qed
\end{pf}

The above result says that there exists a SAS-encoded attacker strategy if and only if the SAAS contains a positive detected state. This implicitly restricts our solution space to SAS-encoded attackers. The following result further says that, in fact, such a restriction is without loss of generality for the solvability of Problem~\ref{prob1}.  
\begin{theorem}\upshape
Supervisory control system $S/G$ is  IS-detectable, i.e., Problem~\ref{prob1} has a solution,  if and only if there exists a SAS $m\in \mathbb{S}( M^s)$ such that $A_m$ is IS-detectable.
\end{theorem}
\begin{pf}
The sufficiency is straightforward. To see the necessity,  suppose that there exists an  IS-detectable
attacker $A$, which may not be encoded as a SAS. 
Let $\alpha\sigma\in P(\mathcal{L}(S_A/G))$ be a shortest sequence such that 
(i) $A$ is stealthy along $\alpha$; and 
(ii) $\mathcal{E}_{S_{A}/G}^I(\alpha\sigma)\subseteq X_{sec}$. 
Therefore, $(\alpha\sigma)_A$ is a well-defined extended string in $M^s$. 
Furthermore, by Proposition~\ref{prop:2}, we have $\mathcal{E}_{S_A/G}^I(\alpha)=I(\tilde{q})$, 
where $f^s((\alpha\sigma)_A)=(q,\tilde{q},z)=q_e$. Therefore, $q_e\in  Q_{det}^+$. 
According to Theorem~\ref{them:state-stat}, again, we can extract a SAS-encoded attack strategy from the SAAS $M^s$ in which $Q_{det}^+$ is non-empty.  
\hfill\qed
\end{pf}

The above two theorems suggest immediately an approach for synthesizing an IS-detectable attack strategy:
\begin{itemize}
    \item 
    First, we build the SAAS $M^s$ and check whether or not it contains a positive detected state; 
    \item 
    If so, then we find a SAS $m\in \mathbb{S}(M^s)$  such that a positive detected state is contained, 
    and its induced strategy $A_m$ is the solution.
\end{itemize}
 We illustrate the procedure by the following example. 
 
\begin{myeg}\upshape 
Still, in our running example, the  SAAS $M^s$ has been  shown in Figure~\ref{fig:6}, where  $Q_{det}^+=\{ (X_3,X_5,z_1) \}\neq \emptyset$. 
A possible SAS $m\in \mathbb{S}(M^s)$ is given in Figure~\ref{fig:7}. 
Compared with $M^s$ in Figure~\ref{fig:6}, $m$ in Figure~\ref{fig:7} only has one out-going transition at attack state $a_{01}$. 
Then according to Theorem~\ref{them:state-stat}, $A_m$ is an IS-detectable strategy.   
\begin{figure}
    \centering
    \begin{tikzpicture}
    [QE/.style={rectangle, draw, thick, scale=0.65}, QA/.style={circle, draw, thick, scale=0.6}]
    \tikzmath{
        \gapx = 1.3;
        \gapy = 1.9;
    }
    \node[QE] (S1) at (0*\gapx,0) {$\boldmath{X_1, \tilde{X}_1, z_0}$};
    \node[QE] (S2) at (-1*\gapx,-1*\gapy) {$X_1, \tilde{X}_2, z_0$};
    \node[QE] (S4) at (1*\gapx,-1*\gapy) {$X_3, \tilde{X}_3, z_2$};
    \node[QE, draw=red!70!black] (S5) at (-2*\gapx,-2*\gapy) {$\emptyset, \tilde{X}_4, z_{\textsf{att}}$};
    \node[QE, draw=green!70!black] (S6) at (-1*\gapx,-2*\gapy) {$X_3, \tilde{X}_5, z_1$};
    \node[QE, draw=blue!70!black] (S7) at (0*\gapx,-2*\gapy) {$X_4, \tilde{X}_4, z_2$};
        
    \node[QA] (A1) at (0*\gapx,0-\gapy/2) {$a_{01}$};
    \node[QA] (A2) at (1*\gapx,0*\gapy-\gapy/2) {$a_{02}$};
    \node[QA] (A3) at (-2*\gapx,-1*\gapy-\gapy/2) {$a_{03}$};
    \node[QA] (A4) at (-1*\gapx,-1*\gapy-\gapy/2) {$a_{04}$};
    \node[QA] (A6) at (1*\gapx,-1*\gapy-\gapy/2) {$a_{06}$};
    \node[QA] (A7) at (2*\gapx,-1*\gapy-\gapy/2) {$a_{07}$};
    
    \draw[->, thick] (0*\gapx, \gapy/4) -- (S1.north);
    \draw[->, thick] (S1) -- node[left, scale=0.6]{$b$} (A1);
    \draw[->, thick] (S1) -| node[below left, scale=0.6]{$c$} (A2);
    \draw[->, thick] (A1) -| node[above left, scale=0.6]{$\epsilon$} (S2);
    \draw[->, thick] (A2) -- node[left, scale=0.6]{$c$} (S4);
    \draw[->, thick] (S2) -| node[above left, scale=0.6]{$d$} (A3);
    \draw[->, thick] (S2) -- node[left, scale=0.6]{$c$} (A4);
    \draw[->, thick] (S4) -- node[left, scale=0.6]{$d$} (A6);
    \draw[->, thick] (S4) -| node[below left, scale=0.6]{$c$} (A7);
    \draw[->, thick] (A3) -- node[left,scale=0.6]{$d$} (S5);
    \draw[->, thick] (A4) -- node[left, scale=0.6]{$c$} (S6);
    
    \draw[thick] (A6) -- node[above left, scale=0.6]{$d$} +(0,-\gapy/2);
    \draw[->, thick] (A7)  |-node[above left, scale=0.6]{$c$}  (S7);
  
    \end{tikzpicture}
    \caption{Synthesized attack strategy represented by a SAS.}
    \label{fig:7}
\end{figure}
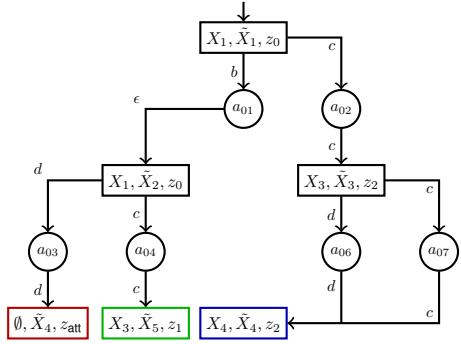
\end{myeg}

\section{Conclusion}\label{sec:conclude}

This paper investigated the  problem of  synthesizing active-sensor attackers against the initial-state opacity of supervisory control systems.  To this end, we defined an information structure, call the AAS,  that embeds all possible attack strategies. Using the structural properties of initial-state estimation, we further simplified the AAS and based on which, an attacker strategy is synthesized. 
Note that, in this work, we only require that the synthesized attacker can \emph{possibly} detected the initial secret along some path.  This is motivated by the negation of opacity, i.e., if the attacker can potentially threaten the system, then the system is not secure. In some cases, we may further require that the synthesized attacker can \emph{always} detected the initial secret along any path.  This synthesis problem with the stronger requirement is currently under investigation. Our preliminary result shows that memory is needed in order to realize such attackers.  
\bibliographystyle{unsrt} 
\bibliography{des} 

\end{document}